# Domain Growth Kinetics in a Cell-sized Liposome

Daisuke SAEKI, Tsutomu HAMADA and Kenichi YOSHIKAWA[*]

Department of Physics, Graduate School of Science, Kyoto University, Kyoto 606-8502

**Abstract**

We investigated the kinetics of domain growth on liposomes consisting of a ternary mixture (unsaturated phospholipid, saturated phospholipid, and cholesterol) by temperature jump. The domain growth process was monitored by fluorescence microscopy, where the growth was mediated by the fusion of domains through the collision. It was found that an average domain size $r$ develops with time $t$ as $r \sim t^{0.15}$, indicating that the power is around a half of the theoretical expectation deduced from a model of Brownian motion on a 2-dimensional membrane. We discuss the mechanism of the experimental scaling behavior by considering the elasticity of the membrane.

---

[*] Corresponding author. E-mail: yoshikaw@scphys.kyoto-u.ac.jp
Tel: +81-75-753-3812    Fax: +81-75-753-3779

**Text**

A raft model has often been adapted during the last decade in order to describe the dynamical feature of cell membrane.[1,2] It was suggested that the mixture of cholesterol and sphingolipids, which are rich in saturated acyl chains, forms dynamic clusters called "rafts", exhibiting Brownian motion within a bilayer membrane.[3,4] Rafts are expected to function as platforms for the attachment of proteins during signal transduction and membrane trafficking.[1] The microdomain structure under the term of raft is regarded as a kind of phase separation that develops due to the interaction between the lipid molecules. In relation to this, a phase separation in monolayers with a different lipid composition at an air-water interface was studied by fluorescence microscopy.[5] Recently, cell-sized lipid vesicles with the diameter on the order of 10 μm have frequently been used as models for living cellular structures (e.g. protein crystallization and gene expression encapsulated inside them),[6-9] and also adapted as a model of raft formation.[10,11] Then, some studies have been performed: the relation between membrane curvature and the domain patterns,[12,13] and the phase diagram in the mixture of lipids and cholesterol.[14-16] As for the domain formation on membrane, theoretical reports on micro phase segregation in cell-sized liposomes have been published.[17,18] Both of these experimental and theoretical studies have focused on domain structures under equilibrium. Contrary to the static stable structure of the phase separation on these studies, it is getting clearer that raft in living cell membranes exhibits significant time-dependent change, including the repetitive cycle of generation and disappearance.[19] Some studies of computer simulation have reported the kinetics of domain growth on liposome.[20,21] Recently, time-dependent change of model membrane has been monitored through direct microscopic observation and scattering experiments.[22-24] However, to our knowledge, no experimental study on the kinetic aspect of micro phase segregation on a cell-sized liposome has appeared yet. To make clear the dynamic property of the raft, such kinetic studies on the phase segregation in simple vesicular system are invaluable. In this study, we conducted a microscopic observation of domain growth in a cell-sized liposome, consisting of ternary mixture of saturated and unsaturated phospholipids along with cholesterol. We measured the time development of the average size of the domains, and found the experimental scaling behavior.

Liposomes were prepared using an unsaturated phospholipid, dioleoyl L-α-phosphatidylcholine (DOPC; Sigma, USA), a saturated phospholipid, dipalmitoyl L-α-phosphatidylcholine (DPPC; Wako, Japan), and cholesterol (Sigma, USA). N-(rhodamine red-X)-1,2-dihexadecanoyl-sn-glycero-3-phosphoethanolamine (rhodamine red-X DHPE, triethylammonium salt; Molecular Probes, USA) was used as a fluorescent probe. Cell-sized liposomes were prepared by a natural swelling method from dry lipid films by the following procedure: 10 mM lipids dissolved in chloroform/methanol (= 2:1) along with 50 mM D(+)-glucose (Nacalai Tesque, Japan) in methanol were dried under vacuum for 2 h to form a thin lipid film. Then, the films were hydrated with distilled water at 60 °C for several hours. The final lipid concentration for liposome was 0.1 mM DOPC, 0.1 mM DPPC, 30 mol% cholesterol, and 0.1 μM rhodamine red-X DHPE, and D(+)-glucose was 0.25 mM. In this three-component bilayer membrane, phase separation between liquid-ordered ($L_o$) and liquid-disordered ($L_d$) phases occurs below the transition temperature (~31 °C).[25] Since the fluorescent probe is localized at $L_o$ phase, two phases are clearly distinguished as bright ($L_o$ phase) and dark ($L_d$ phase) zones by fluorescent microscopy. Liposome solution of 20 μL was placed on a glass plate, and then covered with a slip at a spacing of 100 μm, and sealed. Just before the microscopic observation, we increased the temperature at 60 °C for the period more than 20 s in order to attain a homogenous phase. Then, the samples were cooled down to the room temperature (21 °C), where the time to cool down was ca. 10 s. The process of the domain growth on a liposome was observed by using a confocal laser scanning microscope (LSM 510, Carl Zeiss).

Typical confocal microscopic images of domain growth on a cell-sized liposome are shown in Fig. 1, where two phases are distinguished as bright ($L_o$ phase) and dark ($L_d$ phase) zones. Immediately after the shift of temperature below the miscibility transition, many small domains appear whole over the membrane. The generated domains exhibit random thermal motion, where the smaller ones show larger agitation. The sizes of the domains become larger through the collision and fusion during the thermal motion. Finally, the whole membrane surface was covered with the two different phase regions after several tens of minutes.

We measured the radii of all the visible domains on individual liposome by acquiring 9–11 sliced images along vertical z axis (slice depth < 2.5 μm, interval 3 μm) at every 30 s. Based on the reconstructed 3-dimensional image we have deduced the radius $r$ as the mean value around the domain boundary on individual domains. The time-dependent changes in the average domain radius $r$ on individual liposome during the domain growth process are exemplified in Fig. 2, where $t = 0$ is taken as the time to cross the transition temperature. The growth of the mean domain radius shows a liner relationship in the double-logarithmic plot, where the average of the gradient on seven samples is found to be $0.15 \pm 0.06$, i.e., $r \sim t^{0.15}$.

Next, we discuss the scaling behavior on the domain-growth by considering the effect of the Brownian motion of each domain.[26] The time change of the domain number, $n$, is decreased linearly with frequency of collisions as follows:

$$\frac{dn}{dt} \sim -n\left(\frac{1}{\tau}\right), \qquad (1)$$

where $\tau$ is the mean time in which one domain collides others. Since we are discussing later stage of phase segregation, the total area of domains remains almost constant. This indicates that the number of domains is inversely proportional to the mean domain area, $s$, as $n \sim 1/s$. We introduce a diffusion coefficient of domains moving on the 2-dimensional membrane $D \sim s^{-1/2}$ (by assuming the Stokes-Einstein relation). Substituting a characteristic time $\tau = x^2/4D$ together with the assumption of $x^2 \sim 1/n$ (where $x$ is the mean distance between the domains) into the eq. (1), scaling relationship for the domain radius $r$ on a plain surface ($s \sim r^2$) is deduced as follows:

$$r \sim t^{\frac{1}{3}} \approx t^{0.33}. \qquad (2)$$

Although such simple theoretical argument suggests the existence of the scaling behavior, the expected value of the power, 0.33, is much larger than that, 0.15±0.06, observed in our experiment. Such large discrepancy is probably due to the fact that the above treatment is applicable only in the case of a flat membrane or infinitely large liposome. On the other hand, it has been reported that buckled morphology with large out-of-plane curvatures along the domain boundaries is generated in phase-separated liposomes.[12] We have also confirmed the appearance of the buckled morphology (data not shown). The buckling on the domain is attributable to a competition between the line and bending energies. The line energy between two phases promotes the budding of domains to decrease the boundary length; on the contrary, the bending energy prevents extensive deformation of the membranes. Past numerical studies have suggested that a coupling between the domain patterns and membrane morphologies changes the domain-growth scaling.[20,21] Thus, we will consider the out-of-plane curvature effect in the following.

Regarding that the curvature of the domain portion is approximately proportional to $1/r$, the domain's surface area $s$ is given as $s \sim \frac{\lambda}{\kappa} r^3$, where $\kappa$ and $\lambda$ are the bending rigidity and line tension, respectively.[21] Though the similar argument as in equation (2), we obtain the following scaling law:

$$r \sim t^{\frac{2}{9}} \approx t^{0.22}. \tag{3}$$

It may be obvious that the coupling with the membrane curvature makes the power of domain-growth scaling smaller. Laradji et al., based on numerical simulation with dissipative particle dynamics, reported that the power for total boundary length of all domains, $nr$, was -4/9, corresponding to $nr \sim t^{-\frac{4}{9}}$, which can be transformed form eq. (3).[21] Whereas, our experimental value of 0.15±0.06 is still a little bit smaller than the theoretical value, 0.22. Several additional factors may be considered. It is to be noted that the hydrodynamic friction of domains on a liposome was not described accurately by the above mentioned Stokes relation on an ideal sphere. During the experiments, we noticed the appearance of deformation of the membrane geometry around the two phase boundaries. We should consider the friction involved in the movements of the domains under the effect of out-of-plane curvature in a 2-dimensional fluid membrane.[27] Additionally, we should consider the effect of effective repulsive interaction between neighboring domains (e.g. long range dipolar interaction).[28] It is also to be mentioned that convection on the membrane surface can affect the apparent mobility on the thermal fluctuation. Further experimental studies would be necessary in order to clarify the factors to determine the kinetics of the domain-growth behavior in a cell-sized liposome.

**Acknowledgements**

We thank Dr. M. Ichikawa for his valuable discussion. T. Hamada is supported by a Research Fellowship from the Japan Society for the Promotion of Science for Young Scientists (No. 16000653). This work was supported by the Grant-in-Aid for the 21$^{st}$ Century COE "Center for Diversity and Universality in Physics" and Grant-in-Aid for Scientific Research on Priority Areas (No.17076007) "System Cell Engineering by Multi-scale Manipulation" from the Ministry of Education, Culture, Sports, Science and Technology of Japan.

**Figure captions**

Figure 1. Typical fluorescent microscopic images of domain growth on a cell-sized liposome by using a confocal laser scanning microscope. The times after reaching the miscibility transition temperature are (a) 10 s, (b) 1 min, and (c) 10 min, where numbers of domains on the liposome are (a) 13, (b) 8, and (c) 3, respectively. Domains fused when they collided with each other under thermal fluctuation. Scale bar is 10 μm.

Figure 2. Three examples of double logarithmic plot of the average domain radius $r$ (μm) on individual liposome and time $t$ (s). The slopes are (A) 0.18, (B) 0.15, and (C) 0.17, where the diameters of the liposomes are 24 μm, 19 μm, and 22 μm, respectively.

**Figures**

Figure 1.

(a) 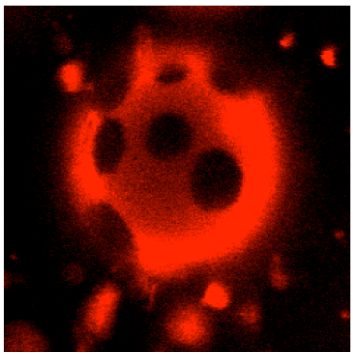  (b) 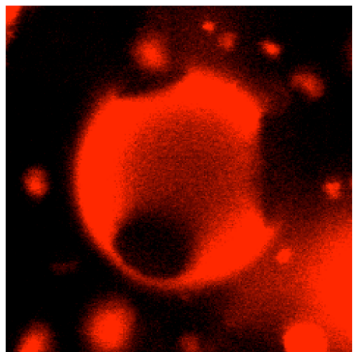  (c) 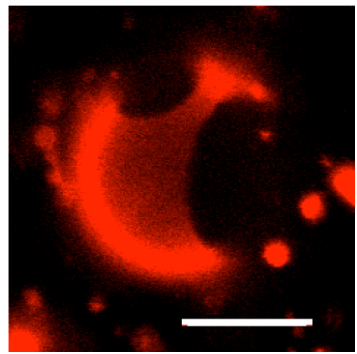

Figure 2.

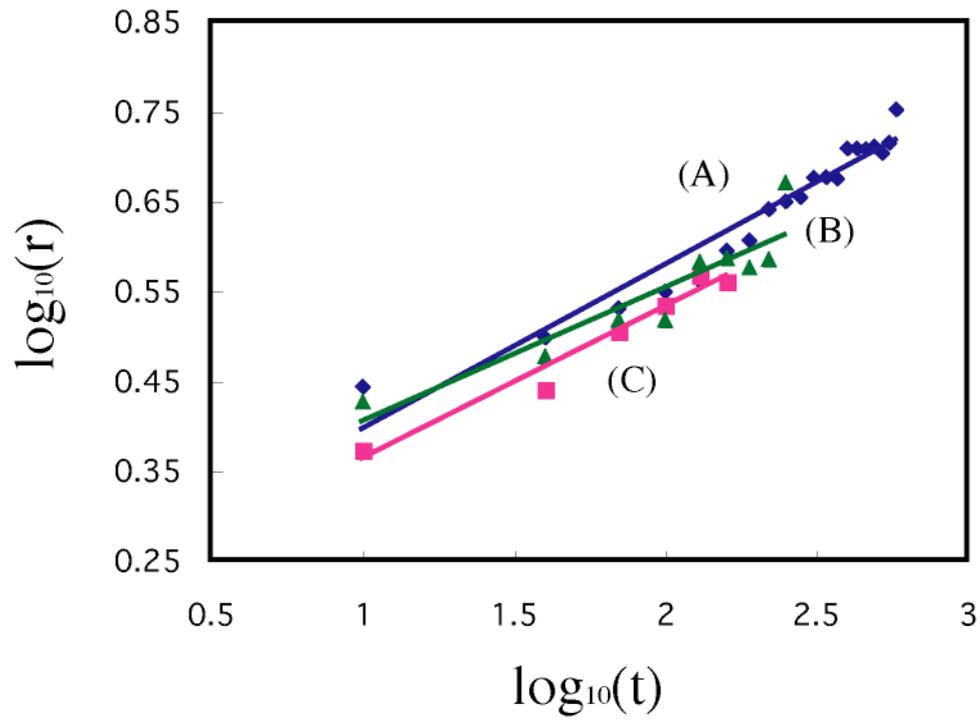